%% file: pinta.tex
\documentclass{pasa}

\usepackage{graphicx}
\usepackage{xspace}
\usepackage[table,xcdraw]{xcolor}
\usepackage{hyperref}
\usepackage{mdframed}
\usepackage{subcaption}
\usepackage{amsmath}
\usepackage{balance}
\usepackage{multirow}

\hypersetup{colorlinks=true,citecolor=blue,linkcolor=blue,urlcolor=blue}

\newcommand{\pinta}{\texttt{pinta}}
\newcommand{\gptool}{\texttt{gptool}}
\newcommand{\rficlean}{\texttt{RFIClean}\xspace}
\newcommand{\filterbank}{\texttt{filterbank}}
\newcommand{\dspsr}{\texttt{dspsr}} 
\newcommand{\timer}{\texttt{Timer}}
\newcommand{\psrchive}{\texttt{PSRCHIVE}}

\newcommand{\gptoolin}{\texttt{gptool.in}}

\newcommand{\wdir}{\texttt{working\_dir}}
\newcommand{\idir}{\texttt{input\_dir}}
\newcommand{\pdir}{\texttt{par\_dir}}
\newcommand{\gdir}{\texttt{gpt\_dir}}
\newcommand{\rcfile}{\texttt{rficconf}}
\newcommand{\cdir}{\texttt{current\_dir}}
\newcommand{\sdir}{\texttt{script\_dir}}

\input{affil.tex}

\title[\pinta{}]{\pinta{}: The uGMRT Data Processing Pipeline for the Indian Pulsar Timing Array}

\author[The InPTA Collaboration]{Abhimanyu Susobhanan$^1$\thanks{\url{s.abhimanyu@tifr.res.in}}, 
Yogesh Maan$^2$, 
Bhal Chandra Joshi$^3$,
T. Prabu$^{4}$,
Shantanu Desai$^5$, 
K. Nobleson$^{6}$,
Sai Chaitanya Susarla$^{7}$,
Raghav Girgaonkar$^{5}$,
Lankeswar Dey$^1$, 
Neelam Dhanda Batra$^8$, 
Yashwant Gupta$^3$,
A. Gopakumar$^1$,
Manjari Bagchi$^{9,10}$,
Avishek Basu$^{3,11}$,
Suryarao Bethapudi$^{12}$,
Arpita Choudhary$^{9}$, 
Kishalay De$^{13}$,
M. A. Krishnakumar$^{14}$,
P. K. Manoharan$^{15}$,
Arun Kumar Naidu$^{16}$,
Dhruv Pathak$^{9,10}$
Jaikhomba Singha$^{17}$,
Mayuresh P. Surnis$^{11}$, 
\affil{$^{1}$\TIFRAstro}%
\affil{$^{2}$\ASTRON}
\affil{$^{3}$\NCRA}
\affil{$^{4}$\RRI}
\affil{$^{5}$\IITHPhys}
\affil{$^{6}$\BITSHydPhys}
\affil{$^{7}$\IISERTvm}
\affil{$^{8}$\IITDPhys}
\affil{$^{9}$\IMSc}
\affil{$^{10}$\HBNI}
\affil{$^{11}$\Jodrell}
\affil{$^{12}$\UTRGVPhysAstro}
\affil{$^{13}$\CaltechCahill}
\affil{$^{14}$\UBielefeldPhys}
\affil{$^{15}$\Arecibo}
\affil{$^{16}$\McGillSpace}
\affil{$^{17}$\IITRPhys}
}

\jid{PASA}
\doi{10.1017/pas.\the\year.xxx}
\jyear{\the\year}

\begin{document}

\begin{frontmatter}
\maketitle

\begin{abstract}
We introduce \pinta{}, a pipeline for reducing the upgraded Giant Metre-wave Radio Telescope (uGMRT) raw pulsar timing data, developed for the Indian Pulsar Timing Array experiment.
We provide a detailed description of the workflow and usage of \pinta{}, as well as its computational performance and RFI mitigation characteristics. 
We also discuss a novel and independent determination of the relative time offsets between the different back-end modes of uGMRT and the interpretation of the uGMRT observation frequency settings, and their agreement with results obtained from engineering tests.
Further, we demonstrate the capability of \pinta{} to generate data products which can produce high-precision TOAs using PSR J1909$-$3744 as an example. 
These results are crucial for performing precision pulsar timing with the uGMRT.
\end{abstract}

\end{frontmatter}

\section{Introduction}
\label{sec:intro}

Ubiquitous galaxy mergers are expected to force their resident supermassive black holes to merge \citep{Berczik2006,Pearson2019}.
During such merger and the preceding inspiral phases, the black hole pairs are expected to emit gravitational waves (GWs) in the nanohertz frequency range \citep{Burke-Spolaor2019,Susobhanan2020}.
Pulsar Timing Arrays \citep[PTAs:][]{HobbsDai2017} aim to detect such GWs by accurately timing the arrival of pulses from an ensemble of millisecond pulsars (MSPs) as these are very precise celestial clocks \citep{Hobbs2020}.
The most promising PTA sources include isolated supermassive black hole binaries (SMBHBs) emitting continuous GWs and an astrophysical stochastic GW background formed from an ensemble of many unresolved SMBHBs   \citep{Burke-Spolaor2019}. 
The rapidly maturing PTA efforts are soon expected to open an additional window to the GW astronomy landscape  inaugurated by the LIGO-Virgo collaboration \citep{LIGOVirgo2019}.

At present, there exist three advanced PTA experiments, namely the Parkes Pulsar Timing Array \citep[PPTA:][]{Hobbs2013,Kerr2020}, the European Pulsar Timing Array \citep[EPTA:][]{Kramer2013,Desvignes2016}, and the North American Nanohertz Observatory for Gravitational Waves  \citep[NANOGrav:][]{McLaughlin2013,Alam2020a,Alam2020b}. 
Additionally, PTA efforts are gaining momentum in India, China and South Africa \citep{Joshi2018_InPTA,Lee2016,Bailes2018}, and these collaborations are referred to as the emerging PTAs. 
The International Pulsar Timing Array (IPTA) consortium combines data and resources from various PTA efforts to enable faster detection of nanohertz GWs \citep{Hobbs2010_IPTA, Perera2019_IPTADR2}. 

The Indian Pulsar Timing Array (InPTA) experiment, operational since 2015 \citep{Joshi2018_InPTA}, aims to use the unique strengths of the Giant Metrewave Radio Telescope  \citep[GMRT:][]{Swarup1991}--- especially after its recent upgrade \citep[uGMRT:][]{Gupta2017}---along with the Ooty Radio Telescope \citep[ORT:][]{Swarup1971,Naidu2015} to complement the other PTA experiments. 
The uGMRT, with its ability to observe below 1 GHz, is an ideal instrument to characterize interstellar medium effects such as dispersion measure (DM) variations of PTA pulsars, which is necessary to achieve the nanosecond timing precision required for the first detection of nanohertz GWs \citep{Joshi2018_InPTA}.

The first step in using uGMRT and ORT data for InPTA science goals is to reduce it to an \textit{archive} format \citep{Hotan2004} -- a pulsar data format widely used among other PTAs. 
Then, this data can be further processed using well-known software to derive various astrophysically relevant quantities including the pulse time of arrival (TOA) and the DM \citep{vanStraten2012}.
This calls for homogeneity in data reduction practices to avoid non-uniformity in the data products used for PTA analysis, which can introduce systematic errors. 
In this paper, we describe a uGMRT pulsar data analysis pipeline named
``\underline{P}ipeline for the \underline{In}dian Pulsar \underline{T}iming \underline{A}rray'' (\pinta{}\footnote{Available at \url{https://github.com/abhisrkckl/pinta}.}), developed for the InPTA experiment to address these concerns as well as to improve the efficiency, reliability, and user friendliness of the data reduction process and to ensure faster turnaround time from observations to PTA analysis.
We have developed \pinta{} with the intention to commission it as a standard pipeline at the GMRT observatory to be used by the wider pulsar community.   
This can help avoid the transfer of large data files by enabling data reduction at the observatory itself. 

For the pipeline to be useful to a wider community, we also discuss how to interpret the uGMRT observation frequency settings. 
{We also present the results of our astronomical experiments carried out to validate the definition of the observing frequency in the engineering specifications of the uGMRT backend hardware and software.} 
Using the same experiment, we also 
{ascertained} the instrumental delays between various back-end modes  used at uGMRT 
{measured through engineering tests}. 
These delays form a crucial piece of information, not only for combining data from multiple bands in the InPTA analysis, but also for other simultaneous multi-frequency observations which use different back-end modes of uGMRT.
 

The outline of this paper is as follows.
A detailed description of the uGMRT raw data as well as the workflow and usage of \pinta{} is provided in Section \ref{sec:pinta_desc}.
Details of the uGMRT observation frequency settings and the astronomical experiments which were used to validate these settings are presented in Section \ref{sec:frequency_settings}.
The performance and RFI mitigation characteristics of \pinta{} are reported in Section \ref{sec:performance}. 
{The ability of \pinta{} to generate data products from which high-precision TOAs can be derived is demonstrated in Section \ref{sec:J1909_timing} using J1909$-$3744 as an example.}
A summary of the \pinta{} pipeline discussed in this paper is given in Section \ref{sec:summary} and our future plans for the development of InPTA-relevant codes including \pinta{} are summarized in Section \ref{sec:future}.

\section{Description of the pipeline}
\label{sec:pinta_desc}

\pinta{} accepts uGMRT raw pulsar timing data 
as input, performs RFI mitigation and folding, and provides the partially folded  pulse profile in the \timer{} archive format \citep{vanStraten2011} as its output.
In what follows, we give a detailed description of the uGMRT raw data and the workflow of the \pinta{} pipeline. 

The thirty GMRT antennas are divided in groups to form multiple subarrays, and each subarray is phased to form voltage beams for two polarizations, 
{and the gains of the two polarizations are equalized during phasing}.
{These voltage beams are then digitized and Fourier transformed (no polyphase filter is employed) to form power spectra across a certain number of frequency channels \citep{Reddy2017}. }
For the phased array (PA) mode that we use in our InPTA timing observations, 
{the spectral powers from the two polarizations are added to form the total intensity $I$ without applying any calibration,} 
and is integrated maintaining the required spectral and time  resolution for the observation  specified in terms of the number of channels $N_{\text{chan}}$ and the sampling time $T_{\text{smpl}}$.
Note that the two polarization voltages can also be combined to 
compute the Stokes parameters \citep[$I$, $Q$, $U$, $V$:][]{Hamakar1996}.
While the recording of the full Stokes data is possible at uGMRT, the implementation of its reduction in the pipeline described here is currently being developed and tested. 
In addition, a real-time coherent dedispersion observing mode is employed to process the voltages to form and record the coherently dedispersed phased array (CDPA) raw data stream \citep{DeGupta2016}. 
Lastly, an incoherent array (IA) data stream can be formed by incoherently adding the spectral powers from different antennas.

The PA and the CDPA total intensity modes are used for  InPTA observations discussed in this paper. 
The CDPA mode is primarily used at the lower frequency bands where the effect of interstellar dispersion is prominent.
The raw data stream from either of these modes, namely a data cube of spectral intensities at $N_{\text{chan}}$ frequency channels for each time sample, are stored as 16-bit integers in a binary raw data file, and the timestamp (in Indian Standard Time) at the start of the observation is saved as a separate ASCII file.
An example timestamp file is shown below.
\begin{mdframed}
\begin{verbatim}
#Start time and date
IST Time: 19:59:57.633098240
Date: 25:08:2018
#Start ACQ SEQ NO = 17
\end{verbatim}
\end{mdframed}
\noindent 
\pinta{} converts the timestamp given in the timestamp file to MJD using \texttt{astropy} \citep{Price-Whelan2018}.
Note that the raw data files do not store any metadata required for downstream processing and it must be provided to the pipeline through a separate file.

Reduction of PTA data involves processing {a large number of such high-volume datasets} (obtained from different MSPs at different epochs in separate bands) through complex processing steps\footnote{{InPTA currently observes six pulsars at bi-weekly cadence simultaneously in two bands. Each such observation creates of the order of 100-150 GB of raw data per band per pulsar.}}. 
In order to ensure that processing can be efficient for such batch processing jobs and to avoid premature run-time failures, a set of checks are done on all the relevant files and folders, and the processing is initiated only if all the checks pass\footnote{These checks include the existence and read/write permissions of the relevant files and folders.}.
If one of the checks fail, an informative error message is shown to enable easier troubleshooting.

\begin{figure*}[t!]
    \centering
    \includegraphics[scale=0.5]{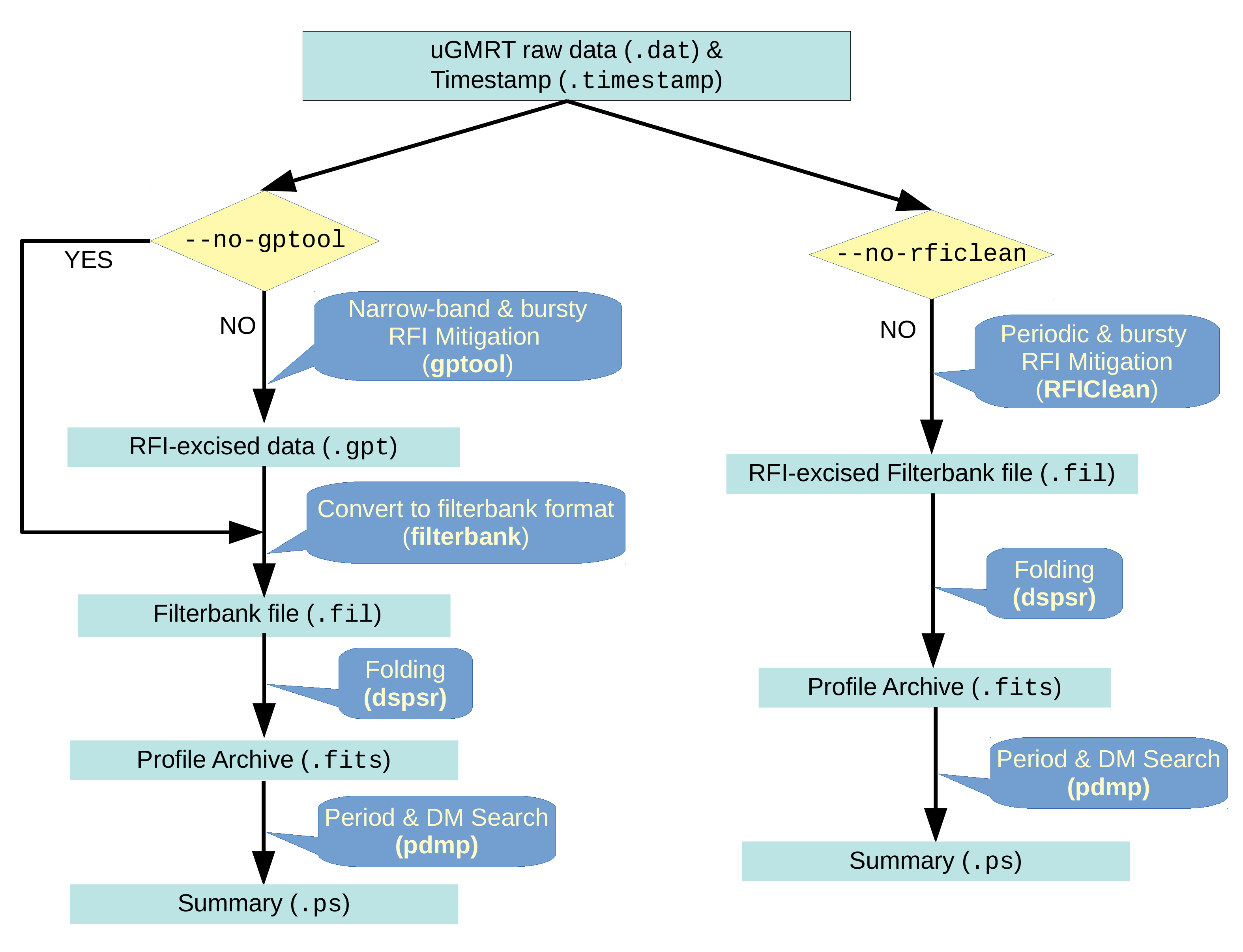}
    \caption{The workflow of \pinta{}.  The pinta pipeline uses uses two separate packages for the RFI
mitigation, namely \gptool{} and \rficlean{}.  A typical data reduction workflow can optionally engage these RFI mitigation choices. 
Note that the profile archives generated by \pinta{} are in the \texttt{Timer} format although their extension is `\texttt{.fits}'. They can be converted to \texttt{PSRFITS} format using \texttt{PSRCHIVE} \citep{Hotan2004}.}
    \label{fig:flowchart}
\end{figure*}

The data processing workflow of \pinta{} is illustrated in Figure \ref{fig:flowchart}. 
\pinta{} uses two separate packages for Radio Frequency Interference (RFI) mitigation, namely \gptool{}\footnote{Stands for GMRT Pulsar Tool} 
\citep{ChowdhuryGupta2021} 
and \rficlean\footnote{Available at \url{https://github.com/ymaan4/rficlean}} \citep{Maan2020}. 
Brief descriptions of these packages are given below.


\subsection{Details of \gptool{}}
\gptool{} is both an RFI mitigation and a data reduction tool for the beamformer data from  GMRT.  
It mitigates both narrow-band spectral line RFI and broadband bursty time-domain RFI.
For narrow-band RFI, it offers a choice of two options for flagging RFI-affected frequency channels: 
(a) it derives a median band-shape and flags channels for which the median absolute deviation (MAD) exceeds a defined threshold;
or (b) it checks for a drop in mean-to-RMS ratio for each channel below a specified threshold to identify channels corrupted by RFI. 
{Our \pinta{} pipeline employs both of these methods available in \gptool{}.}
{For identifying broadband bursty RFI, \gptool{} once again offers two options for removal of outlier time samples, based on different ways of estimating central tendency and variability in the histogram of the frequency-collapsed time series. 
In the first method, a standard median and MAD based scheme is employed to identify RFI-contaminated time samples.
However, when strong RFI is present for a significant duration of the observation time block, the histogram may deviate from unimodality, affecting the robustness of median and MAD estimates.
In such cases, the major mode and the full width at half maximum around the major mode provide robust estimates of the central tendency and variability of the underlying distribution, and a novel scheme for broadband RFI mitigation has been implemented in \gptool{} based on these  statistics.  
This novel scheme has been found to give  superior results, and hence is used in our pipeline.}
For further handling of the channels and time samples that are flagged as RFI by \gptool{}, it  offers two options to the user: either to replace the existing values by zero or to replace the existing values by a local median.
In our pipeline, we use the replace by the local median option as it is known to give better results.
Both the RFI mitigated and unmitigated data can then be dedispersed and folded to the ephemeris of the observed pulsar.
When \gptool{} is run in the interactive mode, the time-series, folded profile and the band-shape are displayed as the tool processes the raw data. 
\pinta{} uses the non-interactive mode of \gptool{}, where the RFI mitigated data, in the same format as the raw input data, is written to an output file along with estimated statistics in  auxiliary files without performing dedispersion or folding.
\gptool{} provides an option for the removal of a baseline computed by dedispersing the data to zero DM,  useful for broadband RFI mitigation, 
and an option for flattening the variations of the band-shape across the observing bandwidth by renormalizing the output of each frequency channel to the same mean value. 
The parameters for RFI removal and the selected modes are specified with a configuration file, named \gptoolin{}.  
\gptool{} has also been extensively used for RFI mitigation in the uGMRT for many other pulsar projects since the beginning of the wide-band observations with the uGMRT \citep{Pleunis2020}.

\subsection{Details of \rficlean{}}
{\rficlean{} excises periodic RFI in the Fourier domain, and then mitigates narrow-band spectral line RFI and broadband bursty time-domain RFI using robust statistics.}
The periodic RFI could severely limit the efficacy of conventional RFI mitigation techniques.
There are many terrestrial sources of periodic interference, the most infamous being the household 50/60\,Hz power-lines. 
\rficlean{} identifies and mitigates  periodic interference in the time series of individual frequency channels using Fourier domain analysis. 
After the excision of periodic interference, \rficlean{} uses the more conventional threshold-based techniques to identify the time samples as well as frequency channels respectively contaminated by broadband bursts and narrow-band RFI.
The identified time samples and frequency channels are replaced by mean values, computed robustly in the local regions around the affected samples. 
\rficlean{} has been extensively and successfully tested against any artefacts which might get incorporated in the data during the periodic RFI excision, and might be relevant to the PTA analysis. 
The details of these tests can be found in \citet{Maan2020}. 
Before inclusion in \pinta{}, \rficlean{} was also independently tested as a stand-alone program using InPTA data, and was found to significantly enhance the quality of the reduced data and the timing analysis. 
For some pulsars with their spin frequency or any of its harmonics unfavourably close to 50\,Hz, detection of the pulsar signal at several epochs was possible only after \rficlean{}'s mitigation of the periodic and other RFI.
\rficlean{} has also been used in several other completed and ongoing projects \citep[e.g.,][]{Maan2019,Oostrum2020}, including in timing experiments and searches for fast radio bursts \cite[][]{SosaFiscella2020,Pastor-Marazuela2020}.

~\\
We note here an important difference between \gptool{} and \rficlean{}: \gptool{} performs band shape normalization on the raw data while \rficlean{} retains the original band shape.
Thus, noticeable difference in shape of the \emph{band-averaged} profiles can occur between the two branches of the pipeline, especially in wide-band observations of pulsars exhibiting significant profile evolution with frequency and interstellar scintillation.
Therefore, we advocate the use of separate templates for generating TOAs from profiles obtained through \gptool{} and \rficlean{}, especially for high precision pulsar timing applications such as PTAs.
{In addition, the use of frequency-dependent two-dimensional templates may also help mitigate this issue \citep{Pennucci2019}}.


\gptool{} accepts uGMRT raw data as input and writes the output in the same format. 
The conversion to the filterbank format is carried out by a version of the \filterbank{} command provided by the \texttt{sigproc} package \citep{Lorimer2011}, customized for uGMRT 
{and distributed along with \pinta{}}.  
On the other hand, \rficlean{} accepts input either in uGMRT raw data format or in the \texttt{sigproc}-filterbank format, and outputs a \texttt{sigproc}-filterbank file.

{It may be illuminating to compare and contrast the RFI mitigation methods available in \pinta{} with that available in the \texttt{CoastGuard} data analysis package\footnote{Available at \url{https://github.com/plazar/coast_guard}} \citep{Lazarus2016} developed for the PSRIX backend of the Effelsberg 100-m Radio Telescope.
\texttt{CoastGuard} provides four algorithms to find and mask or replace channels, sub-integrations and phase bins in the folded profile  contaminated with RFI.
The major difference between the RFI mitigation algorithms available in \pinta{} and \texttt{CoastGuard} is that the former act on raw data whereas the latter acts on folded profile archives.
The mitigation of periodic RFI such as the RFI generated by power distribution lines implemented in \rficlean{} is not possible in the folded profiles.
In addition, the time domain bursty RFI removed by \gptool{} and \rficlean{} typically occur at GMRT at timescales much shorter than our sub-integration interval of 10 s.
These are our main reasons for opting for RFI removal in the raw data rather than folded profiles in our analysis.}

While both the RFI mitigation packages have been well tested, the possibility of discovering  {new artefacts} in the future cannot be ruled out. 
Hence, to avoid the need of re-analysing all the data in such an unlikely future situation, we have designed \pinta{} such that it allows the user to process the data in two separate branches, one for each RFI mitigation package, and produces two separate outputs.  
Availability of data reduced by two independent parts of the pipeline facilitates detailed comparisons and the choice of the optimal RFI mitigation method. 

The RFI-mitigated filterbank files are  folded using \dspsr{} \citep{vanStraten2011} and saved in the \timer{} format, significantly reducing the data volume.
Finally, a period and DM search is performed on the resulting profile archive using the \texttt{pdmp} command provided by \texttt{psrchive}, producing a summary document in the {postscript} format. 
This file is used as a visual check to ensure that the pulsar has been detected and that the analysis has finished successfully.


\subsection{Usage}

\begin{table*}[t]
\begin{tabular}{|l|l|l|}
\hline
\textbf{Argument} & \textbf{Description}  & \begin{tabular}[c]{@{}l@{}}\textbf{Mandatory}/\\\textbf{Optional}\end{tabular} \\ \hline
\multicolumn{3}{|l|}{\textit{\textbf{Positional Arguments}}}\\ \hline

\texttt{\textless{}input\_dir\textgreater{}} & The input directory. & Mandatory \\ \hline
\texttt{\textless{}working\_dir\textgreater{}} & The working directory. & Mandatory \\ \hline
\multicolumn{3}{|l|}{\textit{\textbf{Options}}} \\ \hline
\texttt{-{}-help} & Output a help message. & Optional \\ \hline
\texttt{-{}-test} & \begin{tabular}[c]{@{}l@{}}Do not execute data processing commands. All checks are performed on the \\ input files and the commands are printed on the screen. This option is present\\ for troubleshooting.\end{tabular} & Optional         \\ \hline
\texttt{-{}-no-gptool} & Do not run \gptool{}. Produces an output file without RFI mitigation. & Optional \\ \hline
\texttt{-{}-no-rficlean} & Do not run \rficlean{}. & Optional \\ \hline
\texttt{-{}-nodel} & \begin{tabular}[c]{@{}l@{}}The pipeline deletes all intermediate output files by default to conserve disk\\space. This option preserves the intermediate outputs.\end{tabular} & Optional \\ \hline
\texttt{-{}-retain-aux}  & \begin{tabular}[c]{@{}l@{}}Components of the pipeline produce various side products in addition to the\\ primary data products, which are removed by \pinta{} by default. This option\\preserves these files by moving them to a folder named \texttt{aux} inside the\\\wdir.\end{tabular} & Optional \\ \hline
\texttt{-{}-log-to-file} & \begin{tabular}[c]{@{}l@{}}This option redirects the standard output generated from \pinta{} to a log file\\in the \cdir{}.\end{tabular} & Optional \\ \hline
\texttt{-{}-gptdir \textless{}...\textgreater{}} & \begin{tabular}[c]{@{}l@{}}Specifies the directory where the gptool configuration files are stored.\\ By default, this is specified in the configuration file (See subsection \ref{sec:pinta.yaml}). \end{tabular} & Optional \\ \hline
\texttt{-{}-pardir \textless{}...\textgreater{}} & \begin{tabular}[c]{@{}l@{}}Specifies the directory where the pulsar ephemeris files are stored.\\ By default, this is specified in the configuration file (See subsection \ref{sec:pinta.yaml}).\end{tabular} & Optional \\ \hline
\texttt{-{}-rficconf \textless{}...\textgreater{}} & \begin{tabular}[c]{@{}l@{}}Specifies the \rficlean{} configuration file. By default, this is specified in the
\\  configuration file (See subsection \ref{sec:pinta.yaml}).\end{tabular} & Optional \\ \hline
\end{tabular}
\caption{Command line options available in \pinta{}.}
\label{tab:cmd_options}
\end{table*}

The \pinta{} pipeline can be invoked from the command line with the following syntax.
\begin{verbatim}
  $ pinta [--help] [--test] [--no-gptool]
  [--no-rficlean] [--nodel] [--retain-aux] 
  [--log-to-file] [--gptdir <...>] 
  [--pardir <...>] [--rficconf <...>] 
  <input_dir> <working_dir>
\end{verbatim}
\pinta{} requires specifying two mandatory parameters and a few other optional parameters as inputs as listed below.
\begin{enumerate}
    \item \textbf{Input directory} (\idir) --- The directory where the raw data files and the corresponding timestamp files are stored.
    \item \textbf{Working directory} (\wdir) --- The output files, as well as all the intermediate products, will be written to this directory. This directory must contain a file named \texttt{pipeline.in} as specified in subsection \ref{sec:pipeline_in}, and the user must have `read' and `write' permissions for this directory. The working directory can be the same as the input directory.
    \item \gptool{} \textbf{configuration directory} (\gdir) --- This directory should contain the configuration files required to run \gptool{}, named \texttt{gptool.in.xxx} where `\texttt{xxx}' represents the local oscillator frequency of the uGMRT band.
    \item \textbf{Pulsar ephemeris directory} (\pdir) --- This directory should contain the pulsar ephemeris (\texttt{.par}) files in the \texttt{tempo2} format, required for folding the data. Each ephemeris file should be named \texttt{JNAME.par} where ``JNAME'' is the name of the pulsar in the J2000 epoch.
    \item \rficlean{} \textbf{configuration file} (\rcfile) --- This file contains the settings and flags required to run \rficlean{} for \pinta{}.
\end{enumerate}
In addition, we shall refer to the directory from which \pinta{} is invoked and the directory where the \pinta{} script is stored as the \textit{current directory} (\cdir) and \textit{script directory} (\sdir) respectively.

Note that both \wdir{} and the \cdir{} require write access. 
The \idir{} and \wdir{} are mandatory positional arguments to be passed to \pinta{}, while \gdir{}, \pdir{} and \rcfile{} are by default read from a configuration file, detailed in the next subsection. 
\gdir{}, \pdir{} and \rcfile{} can be explicitly specified in the command line through the \texttt{-{}-gptdir}, \texttt{-{}-pardir} and \texttt{-{}-rficconf} options respectively.
The various options and command line arguments are summarized in Table \ref{tab:cmd_options}. 

\begin{table*}[t]
\begin{tabular}{|l|l|l|l|l|}
\hline
\textbf{Column} & \textbf{Parameter}     & \textbf{Description}                                                                                                                                        & \textbf{Data Type} & \textbf{Unit} \\ \hline
1      & JName         & The name of the pulsar in J2000 epoch.                                                                                                             & String    &      \\ \hline
2      & RawDataFile   & \begin{tabular}[c]{@{}l@{}}Raw data file name. Only the file name is required and \\not the full path.\end{tabular}                                & String    &      \\ \hline
3      & TimestampFile & \begin{tabular}[c]{@{}l@{}}Timestamp file name. Only the file name is required and \\ not the full path.\end{tabular}                               & String    &      \\ \hline
4      & Frequency ($F_{\text{LO}}$)     & Local oscillator frequency of the observing band.                                                                                                & Float     & MHz  \\ \hline
5      & NBins ($N_{\text{bin}}$)        & Number of phase bins for the folded profile.                                                                                                       & Integer   &      \\ \hline
6      & NChans ($N_{\text{chan}}$)       & Number of frequency channels.                                         & Integer   &      \\ \hline
7      & BandWidth ($\Delta F$)  & \begin{tabular}[c]{@{}l@{}}Bandwidth of the observing band.\end{tabular}                         & Float     & MHz  \\ \hline
8      & TSample ($T_{\text{smpl}}$)      & The sampling time used for observation.                                                                                                            & Float     & s    \\ \hline
9      & SideBand      & \begin{tabular}[c]{@{}l@{}}The side-band. This should be either LSB (lower side-band) \\or USB (upper side-band).\end{tabular}                        & String    &      \\ \hline
10     & NPol ($N_\text{pol}$)         & Number of polarizations (1:=(I), 4:=(I,Q,U,V))                                                                                           & Integer   &      \\ \hline
11     & TSubInt ($T_\text{subint}$)      & \begin{tabular}[c]{@{}l@{}}The duration of individual sub-integrations within which\\ the data will be folded over the pulsar period.\end{tabular} & Float     & s    \\ \hline
12     & Cohded        & \begin{tabular}[c]{@{}l@{}}Whether the data has been coherently dedispersed \\ \citep{DeGupta2016}. 1 represents Yes and 0 represents No.\end{tabular}                 & Boolean   &      \\ \hline
\end{tabular}
\caption{Description of various columns in the \texttt{pipeline.in} file.}
\label{tab:pipeline.in_cols}
\end{table*}

\subsection{The Configuration File}
\label{sec:pinta.yaml}

The \pinta{} configuration file stores the default settings required to run the pipeline, such as the \gdir{}, \pdir{} and \rcfile{} in \texttt{YAML} format\footnote{\url{https://yaml.org/}}. 
This file should be named \texttt{pinta.yaml} and stored in the \sdir.

A sample configuration file is shown below.\\

\begin{mdframed}
\begin{verbatim}
pinta:
  pardir: /path/to/pulsar/ephemeris/dir/
  gptdir: /path/to/gptool/config/dir/
  rficconf: /path/to/rfiClean/config/file/ 
\end{verbatim}
\end{mdframed}

\subsection{The \texttt{pipeline.in} File}
\label{sec:pipeline_in}
Since the raw input data files do not contain any metadata required for downstream processing, such as the number of channels and the bandwidth, it must be provided separately. 
\pinta{} accepts this information through a space-separated ASCII file named \texttt{pipeline.in} stored in the \wdir{}.
Each row in \texttt{pipeline.in} corresponds to one raw data file and the various columns are described in Table \ref{tab:pipeline.in_cols}.
Rows starting with ``\#'' are treated as comments and ignored.
\pinta{} processes rows in the \texttt{pipeline.in} files serially until all rows are processed successfully or a validation criterion is not met.

An example \texttt{pipeline.in} file is shown in Figure \ref{fig:pipeline.in_eg}.

\begin{figure*}[t!]
    \centering
    {\tiny
    \begin{mdframed}
\begin{verbatim}
#JName      RawData                         Timestamp                             Freq  Nbin  NChan  BandWidth    TSmpl       SB   NPol  TSubint  Cohded
J1939+2134  J1939+2134.25032019.B3.cdp.dat  J1939+2134.25032019.B3.cdp.timestamp  500   128   1024   100          0.00008192  LSB  1     10.0     1
J1939+2134  J1939+2134.25032019.B4.pa.raw   J1939+2134.25032019.B4.pa.hdr         750   128   1024   100          0.00008192  LSB  1     10.0     0
J1939+2134  J1939+2134.25032019.B5.cdp.dat  J1939+2134.25032019.B5.cdp.timestamp  1460  128   1024   100          0.00008192  LSB  1     10.0     1
\end{verbatim} 
    \end{mdframed}}
    \caption{An example \texttt{pipeline.in} file}
    \label{fig:pipeline.in_eg}
\end{figure*}

\subsection{Storage Requirements}
The uGMRT raw data file generated by an hour-long observation is typically of the order of a hundred  Gigabytes. 
A uGMRT raw data file contains, for each time sample, $N_\text{pol}$ polarization intensities/correlations in $N_{\text{chan}}$ frequency channels represented as 16-bit integers.
In general, the file size of the raw data file for an observation duration  $T_{\text{obs}}$ and sampling time $T_{\text{smpl}}$ is given by
\begin{equation}
    S_{\text{raw}}=N_{\text{pol}}N_{\text{chan}}\frac{T_{\text{obs}}}{T_{\text{smpl}}}\times2\text{ Bytes}\,.
\end{equation}
 
The intermediate products generated by the pipeline, namely, \texttt{.gpt} and \texttt{.fil} files, will have roughly the same size as the input file along with a small header which stores observation metadata. 
The output archive files are typically smaller, of the order of hundreds of Megabytes in size, since we fold the raw data over longer sub-integrations. 
The size of the output archive, excluding the header, is approximately given by
\begin{equation}
    S_{\text{arch}} \sim \frac{T_{\text{smpl}}}{T_{\text{subint}}} N_{\text{bin}} S_{\text{raw}}\,,
\end{equation}
where $T_{\text{subint}}$ is the duration of a sub-integration and $N_{\text{bin}}$ is the number of phase bins in the profile.
{In our analysis, we typically use $T_{\text{subint}}=10$ s.}
In general, the maximum amount of disk space required by \pinta{} is less than four times the total size of the raw data files, while preserving all intermediate files (i.e., using the \texttt{-{}-nodel} option). 
If the \texttt{-{}-nodel} option is not used, the maximum amount of disk space required is approximately the size of the largest raw data file.

\section{Interpretation of observatory frequency settings }
\label{sec:frequency_settings}

The GMRT Wide-band Back-end \citep[GWB; ][]{Reddy2017} provides three different observation modes, namely IA, PA or CDPA, as described in Section \ref{sec:pinta_desc}.
The settings used during a pulsar observation depend on the band of observation and the mode of the observatory back-end. 
These settings are required for data reduction using \pinta{} and are communicated to the pipeline through a \texttt{pipeline.in} file as mentioned in Section \ref{sec:pipeline_in}. 
As the frequency labelling of the pulsar data cube varies with the back-end mode used, these need to be determined and encoded in \pinta{} in a manner which simplifies the specification of observation settings for the user. 

The times of arrival (TOAs) of a pulsar pulse recorded simultaneously in two bands A and B, using back-end modes P and Q respectively, are related by 
\begin{equation}
    t_{AP}-t_{BQ} = \Delta_{PQ} + \mathcal{D}\times \text{DM}\left( F_{1A}^{-2} - F_{1B}^{-2} \right)\,,
    \label{eq:mode_delay}
\end{equation}
where $t_{AP}$ and $t_{BQ}$ are the TOAs, $\Delta_{PQ}$ is the relative instrumental offset between modes $P$ and $Q$, $\mathcal{D}$ is the dispersion measure constant, $\text{DM}$ is the dispersion measure of the pulsar at the epoch of observation, and $F_{1A}$ and $F_{1B}$ are the frequency labels of the channels to which the signals in bands A and B are dedispersed.
Both the offsets $\Delta_{PQ}$ and the frequency labels $F_{1X}$ (where $X$ represents the band of observation) are crucial for performing precision pulsar timing using uGMRT. 
These are defined as part of the engineering specifications of the GWB hardware and software \citep{Reddy2017,DeGupta2016}. 
Engineering tests with standard inputs to the hardware were carried out to verify these definitions 
{and revealed that there is no offset between  time series in IA and PA mode, whereas a 1 buffer (256 Mbytes) offset exists between IA/PA and CDPA modes. 
This offset is 0.67108864 s for 200 and 400 MHz bandwidths and 1.34217728 s for 100 MHz bandwidth, and this was verified up to 5 ns precision in engineering tests. 
Likewise, the frequency definitions were worked out from engineering considerations and tested in an engineering sense with fixed frequency tones. 
While the precision of astronomical tests is not likely to be high due to system noise and coarser sampling, nevertheless such tests with wide-band radio emission are also needed to gain confidence, particularly for coherently dedispersed data.}
In this section, we describe the astronomical tests carried out to validate the frequency labeling $F_{1X}$
to be encoded in \pinta{}, and to determine the offsets $\Delta_{PQ}$.

\subsection{Calibration experiment}
\label{sec:frequency:expt}

\newcommand{\dmunit}{pc\,cm$^{-3}$}

\begin{figure*}
\begin{subfigure}{.5\textwidth}
        \centering
        \includegraphics[scale=0.35]{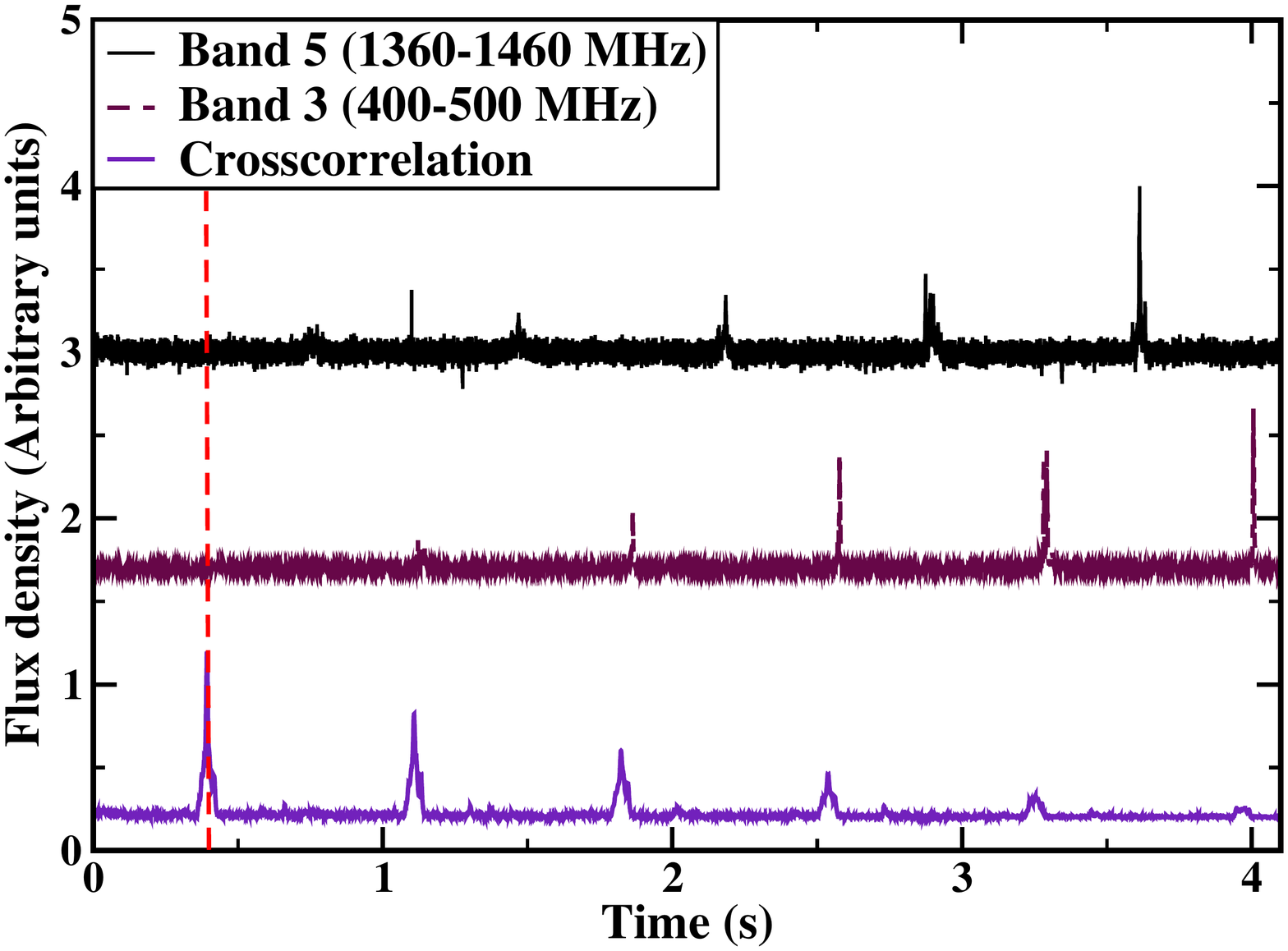}
        \caption{PSR J0332+5434}
        \label{fig:freqa}
\end{subfigure}
\begin{subfigure}{.5\textwidth}
        \centering
        \includegraphics[scale=0.35]{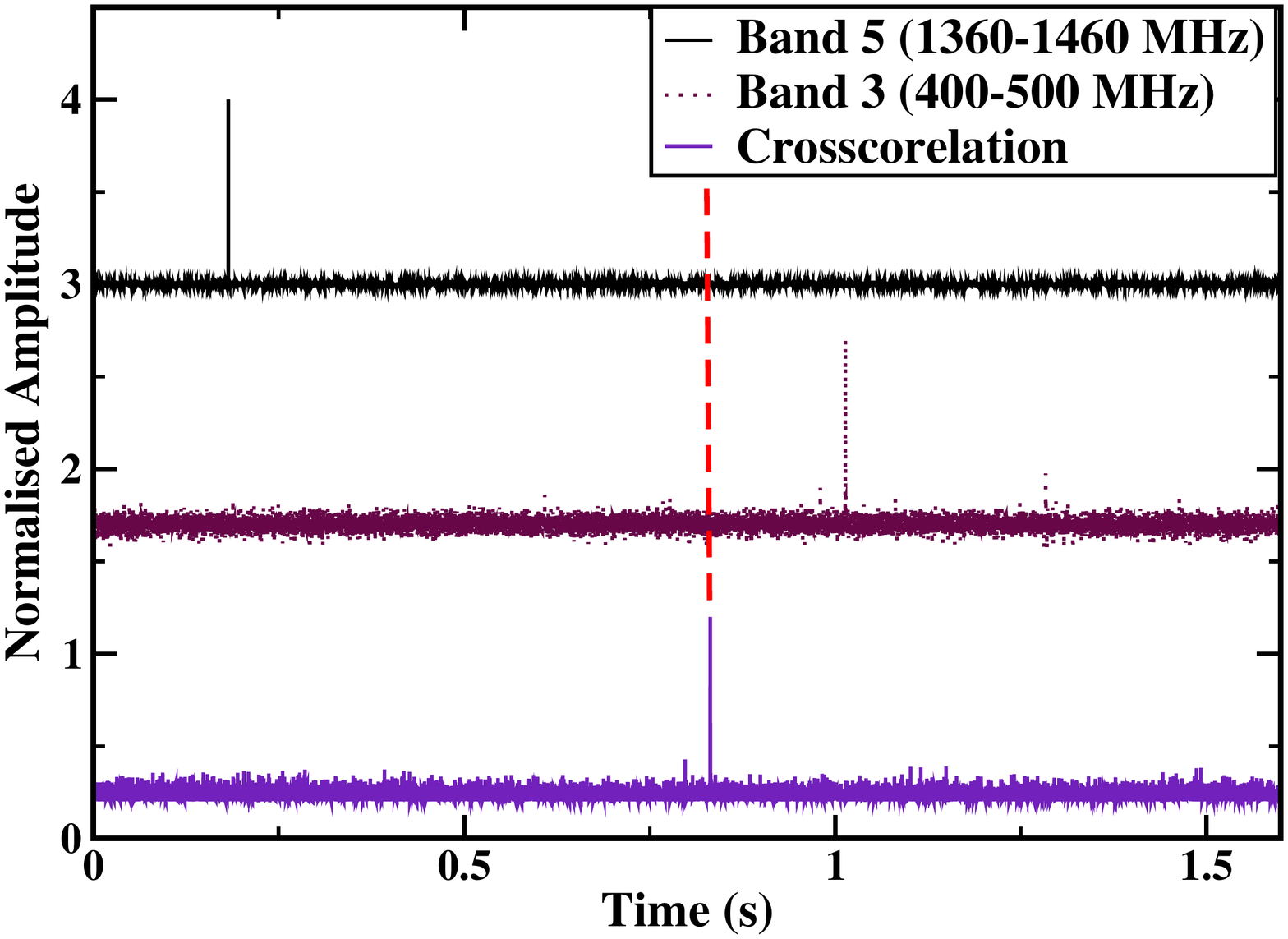}
        \caption{PSR J0534+2200}
        \label{fig:freqb}
\end{subfigure}
\label{figfreq}
\caption{Time series observed using Band 5 (1360 -- 1460 MHz : top plot in each panel) and Band 3 (400 -- 500 MHz : middle plot in each panel) was used to determine the delay between the two bands using pulsars PSRs J0332+5434 and J0534+2200. The delay is obtained from the lag measured using the cross-correlation (shown in the bottom plot of each panel) of the two time series. 
The delay in each case was compared with that expected {(labeled with vertical red dashed lines in the plot)} due to dispersion in ionised interstellar medium to determine both the frequency definition as well as relative pipeline delays : 
(a) Observations of single pulses of the bright pulsar J0332+5434 showing a delayed single pulse pattern in Band 3 compared to Band 4, 
(b) Observations of a Giant pulse of  PSR J0534+2200 where the delay between Band 5 (top plot) and Band 3 (middle plot) was found consistent with that expected due to dispersion, assuming the correct frequency definitions (Equations \ref{eq:Fpa} \& \ref{eq:Fcdp}) and zero relative fixed pipeline delay.}
\end{figure*}

The required frequency labeling and the instrumental offsets were validated using observations of the Crab pulsar (PSR J0534+2200) and PSR J0332+5434. 
The former is a bright pulsar with 33.7 ms period and a relatively high DM  \citep[56.7 \dmunit{} :][]{Lyne2014}. 
The DM of the Crab pulsar varies from epoch to epoch and this pulsar exhibits sporadic intense pulses, called giant pulses \citep[GPs; ][]{lcu+95,hkw+03}, typically once every four minutes 
{at uGMRT frequencies at uGMRT sensitivity}. 
The GPs provide a time marker, which is a strong function of frequency due to  interstellar dispersion. 
Moreover, the arrival times of this marker across different frequencies vary with epoch due to DM variations. 
Thus, GPs provide a sensitive probe to validate the assumed frequency labels for the spectral data. 
{PSR J0332+5434, with a flux density of $\sim$1500 mJy at 408 MHz, is the brightest pulsar in the northern hemisphere at metre-centimetre wavelengths with a period of 714 ms and a DM of 26.76  pc\,cm$^{-3}$ \citep{Lorimer1995,Hassal2012}. }
Bright single pulses with pulse-to-pulse intensity variations interspersed with pulse nulls are seen in this pulsar (see Figure \ref{fig:freqa}). 

The GWB can simultaneously be used in its different modes of operation in different bands using any combination of the four beams provided \citep{Gupta2017,Reddy2017}. 
This capability was exploited to record data on GPs from the Crab pulsar  and single pulses from PSR J0332+5434 in IA, PA, and CDPA modes of GWB using different frequency bands available with the uGMRT. 
{For the Crab pulsar, first the GPs were identified in IA, PA and CDPA mode data at both Band 3 and Band 5. 
We investigated the cross-correlation in the recorded time series around the identified GPs from different modes and frequency-bands to determine the lag in the arrival times of the GPs. } 
This lag, recorded for example with PA in Band 5 and CDPA in Band 3, depends on the DM of the pulsar (specified up to a precision of 0.001  pc\,cm$^{-3}$) and the frequency labeling used for the two bands, as given by equation \ref{eq:mode_delay}.
As the DM time series of this pulsar is known to the required precision from independent measurements \citep{Lyne1993,Lyne2014} made public by the Jodrell Bank Observatory\footnote{\url{http://www.jb.man.ac.uk/pulsar/crab/crab2.txt}}, 
the expected lag in the arrival times of identified GPs was calculated from the DM nearest to the epoch of observations. 
Hence, any difference between the expected and measured lags is due to either (a) incorrect frequency labeling, or (b) relative time offset between the two modes. 
As the DM of this pulsar varies over a timescale of one month, two observations separated by one month will yield different delays due to frequency labeling, whereas the relative instrumental delay is expected to be constant.
Thus, both the frequency labeling as well as relative offsets can be simultaneously determined by two such observations. 
We check these results for consistency using similar analysis with PSR J0332+5434. 

\subsection{Calibration observations and results}
\label{sec:frequency:calobsres}
\begin{table*}[t]
  \begin{center}
    \begin{tabular}{|l|c|l|c|c|c|c|}
      \hline
      \textbf{Epoch} & \textbf{DM} & \textbf{Bands and Modes} & \textbf{Sampling Time} & \textbf{Expected } &\textbf{Observed } & \textbf{$\Delta_{PQ}$}\\
      \textbf{(MJD)} & \textbf{(pc\,~cm$^{-3}$)} &  & \textbf{($\mu$s)} & \textbf{delay (s)} &\textbf{delay (s)} & \textbf{s} \\ \hline
      \multirow{4}{*}{58832} & \multirow{4}{*}{56.7528} & B5CDPA--B3CDPA & 81.92 & 0.83172 & 0.83165(8) & 0.0\\ 
                             &                          & B3CDPA--B5PA   & 81.92 & 0.83173 & 0.83173(8)   & 1.34218 \\
                             &                          & B5CDPA--B5PA  & 81.92 & 0.00001 & 0.00008(8) & 1.34226   \\
                             &                          & B5PA--B3PA   & 81.92 & 0.83137 & 0.83141(8) & 0.0\\
      \hline
      \multirow{3}{*}{58871} & \multirow{3}{*}{56.7401} & B5CDPA--B3CDPA & 20.48 & 0.83259   & 0.83259(2) & 0.0\\ 
                             &                          & B3CDPA--B5PA  & 81.92 & 2.1748 & 2.1746(2) &  1.342177\\
                             &                          & B5PA--B3PA   & 81.92 & 0.8312 & 0.8312(1) & 0.0\\
      \hline
      \multirow{5}{*}{58991} & \multirow{5}{*}{56.7781} & B5CDPA--B3CDPA & 5.12 & 0.8374 & 0.8375(1) & 0.0\\ 
                             &                          & B5CDPA--B4PA  & 40.96 & 0.39062 & 0.39051(8)& 0.0\\
                             &                          & B3CDPA--B4PA   & 40.96 & 0.4468 & 0.4470(2) & 0.0 \\
                             &                          & B4PA--B5PA    & 40.96 & 0.4470 & 0.4472(2) & 0.0\\
                             &                          & B5PA--B5IA    & 40.96 & 0.0 & 0.0 & 0.0\\
      \hline 
    \end{tabular}
    \caption{Results of time delay measurements simultaneously at two different frequency using PSR J0534+2200 for validating frequency definitions and relative pipeline delays ($\Delta_{PQ}$)  for different modes of pulsar observations. The epoch of observations is given in the first column along-with Dispersion measure at that epoch in second column followed by sampling time used, expected and observed delay in samples for different combination of modes at the two frequencies in fourth, fifth, sixth, seventh and third column respectively. The last column presents the relative pipeline delays ($\Delta_{PQ}$). The abbreviations B5CDPA, B3CDPA, B5PA and B3PA indicate data acquisition using Band 5 in CDPA mode, using Band 3 in CDPA mode, Band 5 in PA mode, and Band 3 in PA mode respectively.}
    \label{tab:fd}
  \end{center}
\end{table*}

Calibration observations were carried out on 2019 December 16 (MJD 58832), 2020 January 24 (MJD 58871), and 2020 May 22 (MJD 58991). 
The estimated lags for one combination of modes on 2020 January 24 are shown in Figures  \ref{fig:freqa} and \ref{fig:freqb}. 
The relative offsets and frequency labeling were then determined by matching the measured and expected lags, given by equation (\ref{eq:mode_delay}), and the estimated relative offsets for different modes are tabulated in Table \ref{tab:fd}. 
{While the uncertainty on measurements of these relative pipeline delays ranges from 10 to 80 $\mu$s due to coarser sampling and system noise, these measurements are consistent with the engineering measurements}. 
The relative pipeline delays measured as a result of tests conducted in the first two epochs were corrected in the software by the GMRT engineering team in April 2020. 
This was verified in the tests conducted on May 22, 2020 as can be seen from Table \ref{tab:fd}.


The frequency labeling  $F_{1X}$ for the different modes are expressed in terms of the value of the highest frequency channel  in the following expressions:

\noindent For IA and PA,
\begin{subequations}
\begin{equation}
F_{1X}=\begin{cases}
F_{\text{LO}} & \text{for LSB}\\
F_{\text{LO}}+\Delta F & \text{for USB}
\end{cases}\,,
\label{eq:Fpa}
\end{equation}
and for CDPA,
\begin{equation}
F_{1X}=\begin{cases}
F_{\text{LO}}-\frac{\Delta F}{N_{\text{chan}}} & \text{for LSB}\\
F_{\text{LO}}+\Delta F\left(1-\frac{1}{N_{\text{chan}}}\right) & \text{for USB}
\end{cases}\,.
\label{eq:Fcdp}
\end{equation}
\end{subequations}
Here, $F_{\text{LO}}$ refers to the Local Oscillator (LO) frequency (MHz) used  for the observations, $\Delta F$ is the acquisition bandwidth (typically 100 or 200 MHz) and $N_{\text{chan}}$ denotes the number of  channels or sub-bands across the band. 
The expression is different for each side-band denoted by USB or LSB. When $F_{\text{LO}}$ is chosen at the lowest edge of the band being used, this is called upper side-band (USB) where frequencies are ordered from lowest to highest frequency. The reverse order of frequencies are used in lower side-band (LSB) with  $F_{\text{LO}}$ chosen at the highest edge of the band.
Equations \ref{eq:Fpa}-\ref{eq:Fcdp} are in agreement with what is expected from the implementation of the IA, PA, and CDPA pipelines in GWB \citep{Reddy2017,DeGupta2016}.

These equations are implemented in \pinta{} to make it simpler for the user to use our data reduction pipeline. 
The user specifies the LO frequency, the side-band, the acquisition bandwidth and the number of sub-bands/channels in the \texttt{pipeline.in} file using the same values as specified for the back-end observation setup. 
{The relative offsets determined in these experiments are not coded in \pinta{}, but are included as jumps 
while performing any timing analysis of the uGMRT data.}

\section{Performance}
\label{sec:performance}

To validate the pipeline and investigate its performance, we performed a series of tests using a variety of uGMRT datasets with varying data volume and observation frequencies. 

\begin{figure}[ht]
    \centering
    \includegraphics[scale=0.3]{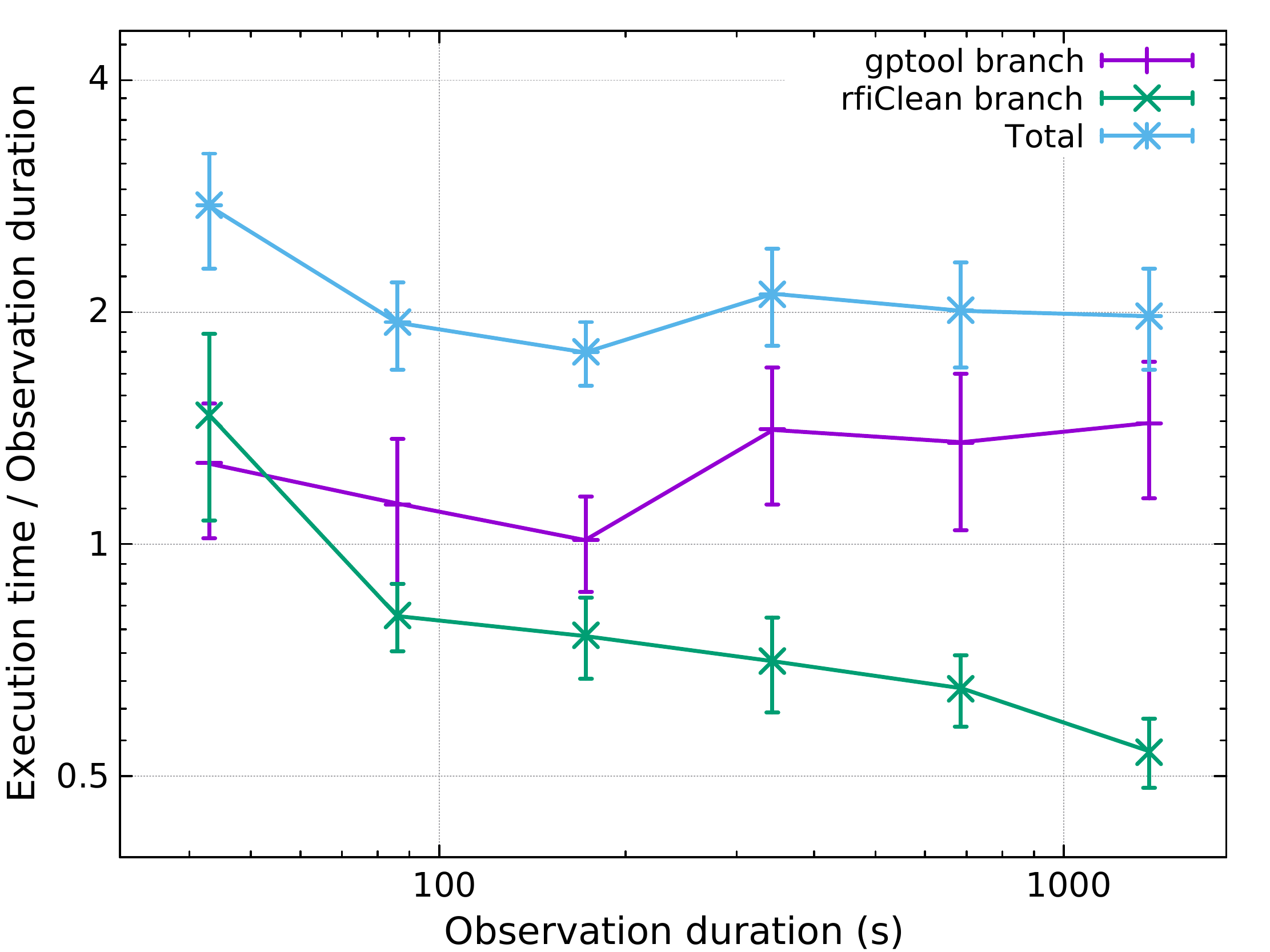}
    \caption{Ratio of execution time by observation duration (observe-to-reduce ratio) plotted versus the observation duration.
    The observe-to-reduce ratio for each of the two branches of \pinta{} as well as the same for the entire pipeline is plotted.
    Each data point represents the median of 10 tests and the error bars represent the corresponding median absolute deviation. }
    \label{fig:exectime_ratio} 
\end{figure}

{To gauge the computational performance of \pinta{}}, we sliced the raw data files from ten different observations (the details of these datasets are given in Table \ref{tab:test_datasets}) into file sizes of 1 GiB\footnote{1 GiB = $2^{30}$ Bytes.}, 2 GiB, 4 GiB, 8 GiB, 16 GiB and 32 GiB, processed each slice separately in \pinta{}, and in each case recorded the execution time of each component of \pinta{} as well as the total execution time. 
The result of this exercise is shown in Figure \ref{fig:exectime_ratio} where 
the ratio of the execution time to the observation duration (observe-to-reduce time ratio) is plotted against the observation duration. 
Each point in Figure \ref{fig:exectime_ratio} represents the median of ten test cases and the error bar represents the corresponding median absolute deviation.
This plot shows the observe-to-reduce ratio to be approximately between ~1.5 and ~3, and that it is not strongly dependent on the data volume.
This behavior is desirable and the observe-to-reduce ratio can indeed be improved to be better than real-time by optimizing and parallelizing the pipeline, which we plan to do in the future.
Such improvements can in principle allow \pinta{} to be deployed as a real-time observatory pipeline for pulsar data reduction.
We also note that the observe-to-reduce ratio while using only one of the two branches is close to or better than real-time.

{To ensure the reliability of the pipeline, these tests were repeated by multiple users on the same datasets mentioned above using different command line options, and the results were compared with each other as well as with results obtained by running the various data reduction codes used in pinta directly to ensure that the results are reproducible.
}

\subsection{RFI Mitigation}

\begin{figure}
    \centering
    \includegraphics[scale=0.4,angle=0]{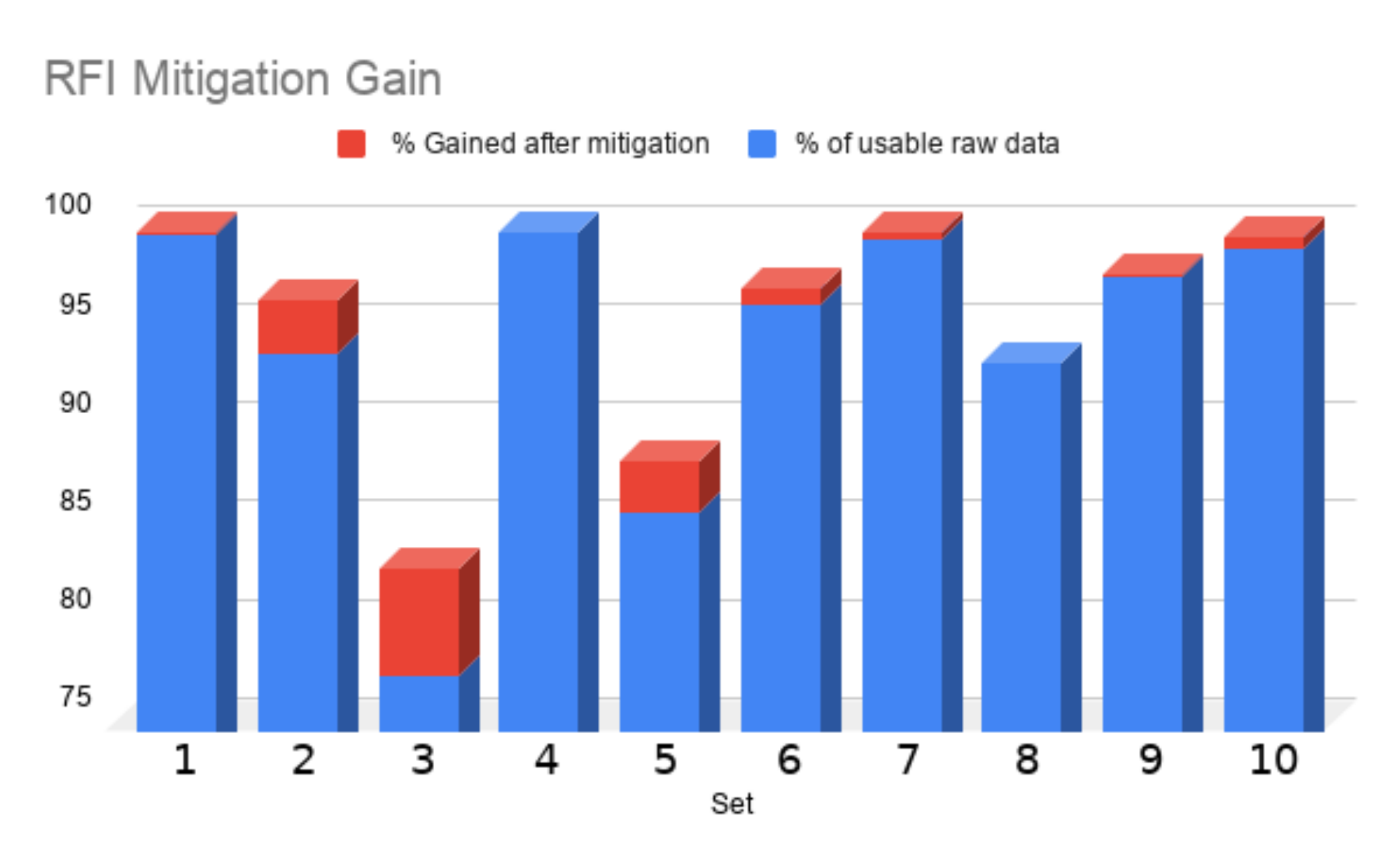}
    \caption{Effectiveness of RFI Mitigation. 
    Each bar represents one dataset.
    The details of each dataset is given in Table \ref{tab:test_datasets}.}
    \label{fig:datagain} 
\end{figure}

RFI mitigation is one of the most important processing steps in the \pinta{} pipeline. 
In order to illustrate the RFI mitigation in the pipeline, we present here a study on ten different datasets (see Table \ref{tab:test_datasets}), each having varying levels of RFI. Data segments were selected from the uGMRT observation bands 3, 4 and 5, MJD 58260-58389 with a total length for the segments 11544 seconds. 
The data quality of each segment prior to and after the \pinta{} RFI mitigation was studied. 
The \texttt{rfifind} command of  \texttt{PRESTO} \citep{Ransom2011} was used to report the percentage of good  intervals in the data. 
The percentage of good intervals that is gained after the RFI mitigation is shown (in red) in  Figure \ref{fig:datagain}. 
{This study provides a feel for the typical RFI mitigation available in the pipeline, and we see from Figure \ref{fig:datagain} that the degree of improvement after applying RFI mitigation varies greatly from dataset to dataset, which is expected since the RFI environment itself is highly variable.
Dataset 3 is of specific interest as the percentage of good intervals more than doubles after applying RFI mitigation, and the pulsar was detected in this dataset only after applying RFI mitigation.}

\begin{table*}[t]
    \centering
    \begin{tabular}{|c|c|c|c|c|}
\hline
\textbf{Dataset} & \textbf{Pulsar} & \textbf{Date} & \textbf{Band} & \textbf{\begin{tabular}[c]{@{}l@{}}Coherent\\ Dedispersion\end{tabular}} \\ \hline
1                & J1857+0943      & 25/08/2018    & 5             & Yes                                                                      \\ \hline
2                & J2145$-$0750      & 22/05/2018    & 4             & No                                                                       \\ \hline
3                & J2145$-$0750      & 25/08/2018    & 3             & Yes                                                                      \\ \hline
4                & J2145$-$0750      & 10/09/2018    & 3             & Yes                                                                      \\ \hline
5                & J1939+2134      & 21/05/2018    & 3             & Yes                                                                      \\ \hline
6                & J1939+2134      & 28/09/2018    & 4             & No                                                                       \\ \hline
7                & J1713+0747      & 07/06/2018    & 5             & Yes                                                                      \\ \hline
8                & J2124$-$3358      & 10/09/2018    & 3             & Yes                                                                      \\ \hline
9                & J1643$-$1224      & 08/07/2018    & 3             & Yes                                                                      \\ \hline
10               & J1643$-$1224      & 25/08/2018    & 5             & Yes                                                                      \\ \hline
\end{tabular}
    \caption{The details of the datasets used for characterizing the performance and RFI mitigation efficacy of \pinta{}.
    Bands 3, 4 and 5 represent 400--500 MHz, 650--750 MHz, and 1360-1460 MHz respectively for our observations.}
    \label{tab:test_datasets}
\end{table*}

\begin{figure*}[t]
\begin{subfigure}{.5\textwidth}
        \centering
        \includegraphics[scale=0.45]{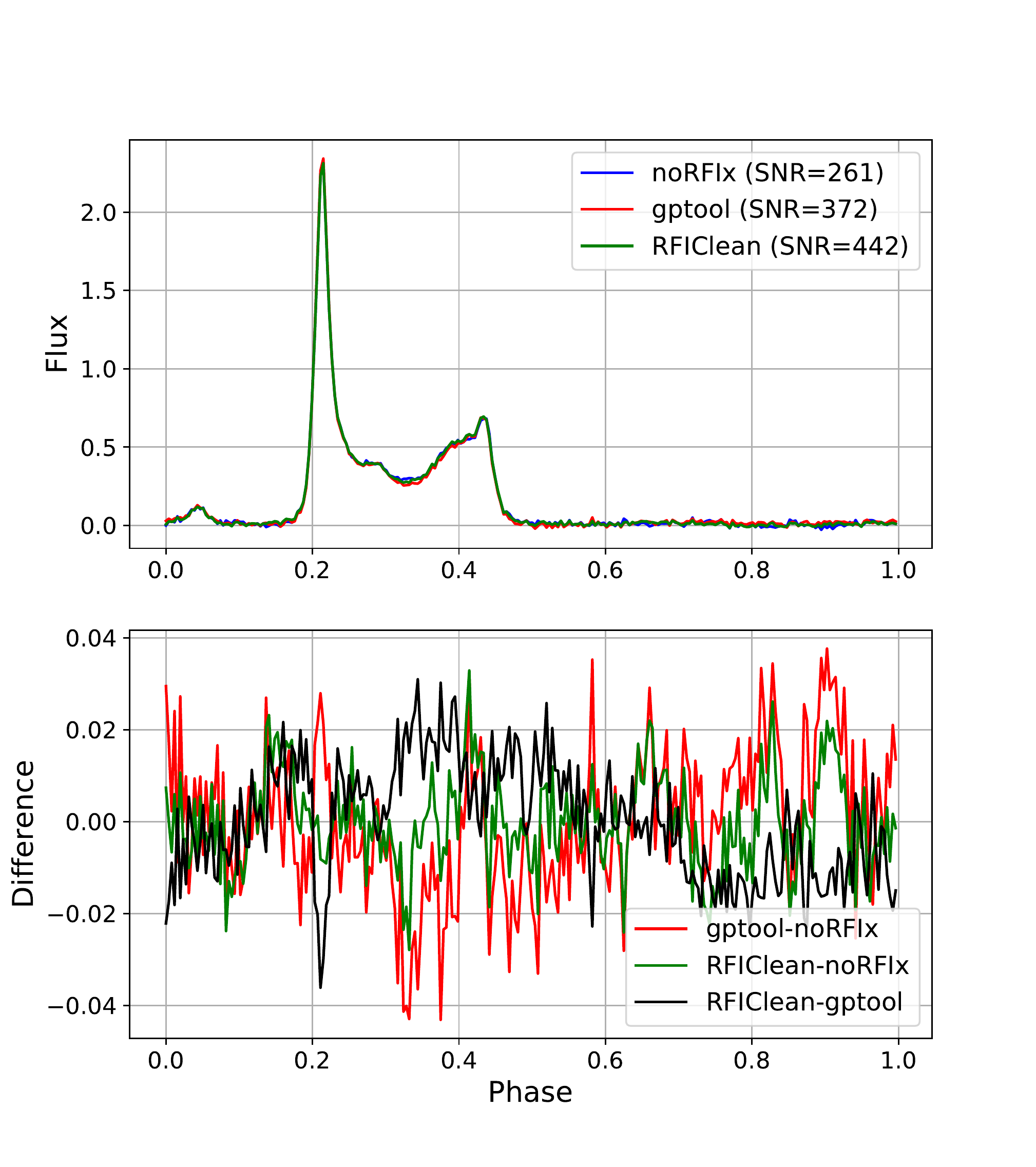}
        \caption{~}
\end{subfigure}
\begin{subfigure}{.5\textwidth}
        \centering
        \includegraphics[scale=0.45]{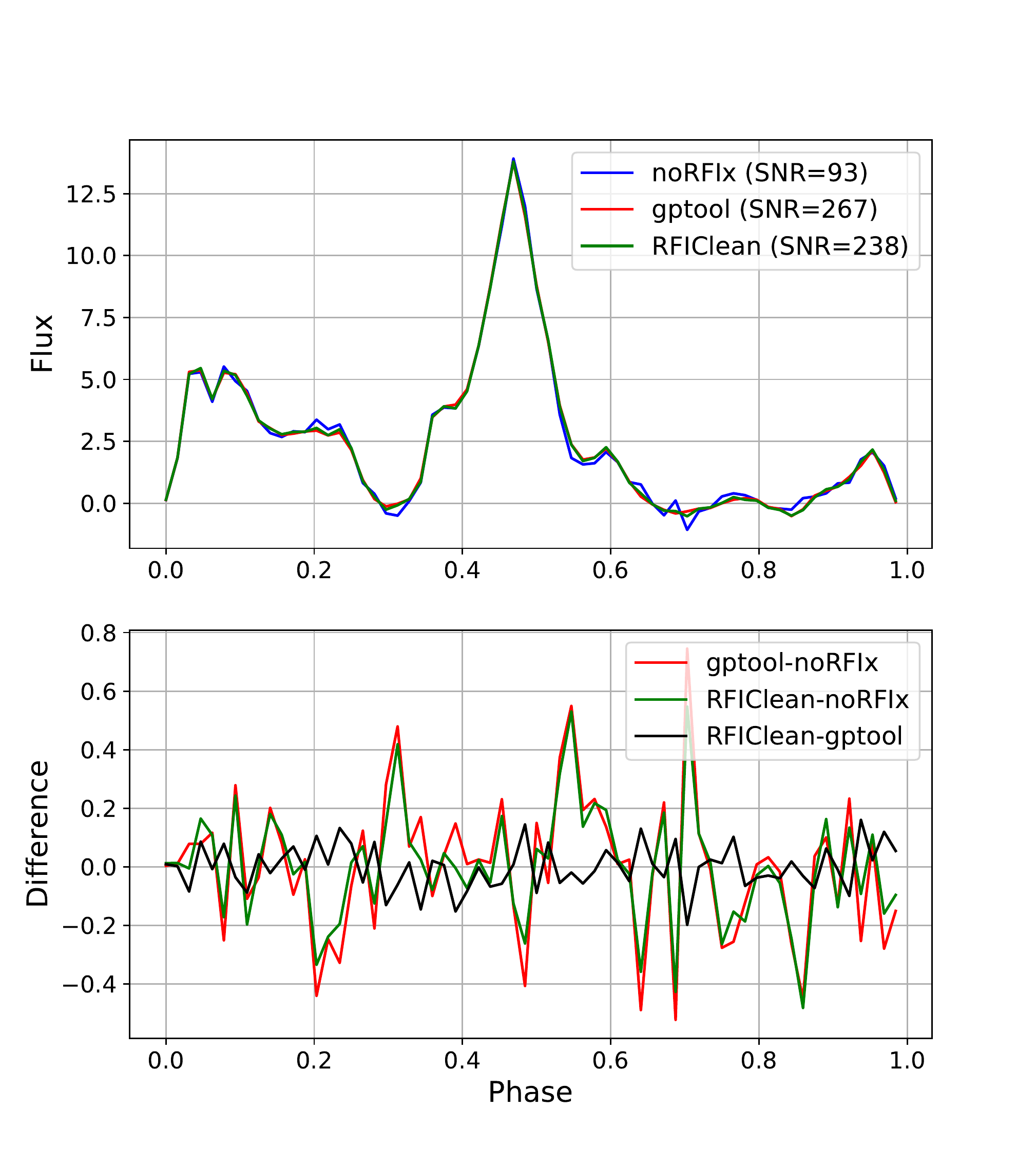}  
        \caption{~}
\end{subfigure}
\caption{Comparison of frequency collapsed profiles obtained using \gptool{}, \rficlean{}, and without any RFI mitigation (\texttt{noRFIx}).
The \texttt{noRFIx} profiles are generated using the \texttt{-{}-no-gptool -{}-no-rficlean} options.
The fluxes are uncalibrated and are in arbitrary units.
The SNRs reported in the plots are obtained using the \texttt{pdmp} command.
Both epochs show significant improvement in SNR while using RFI mitigation.
(a) PSR J2145$-$0750 observed on 16 June 2020 in Band 5 (1260-1460 MHz) with 40.96 $\mu$s sampling time and no coherent dedispersion. The total integration time is 55 min.
(b) PSR J2124-3358 observed on 25 August 2018 in Band 3 (400-500 MHz) with 81.92 $\mu$s sampling time with coherent dedispersion. The total integration time is 24 min.
}
\label{fig:profile_comparison}
\end{figure*}

To further illustrate the efficacy of the RFI mitigation available in \pinta{}, we show  in Figure \ref{fig:profile_comparison} pulse profiles generated using \gptool{}, \rficlean{}, and without performing any RFI mitigation for two observations.
The profiles without any RFI mitigation are produced by running \pinta{} with \texttt{-{}-no-gptool -{}-no-rficlean} options.
{The signal to noise ratios (SNRs) quoted in Figure \ref{fig:profile_comparison} are computed using the \texttt{pdmp}\footnote{\url{http://psrchive.sourceforge.net/manuals/psrstat/algorithms/snr/}} command of \psrchive{}.}
In light of the caveat regarding band shape normalization discussed in Section \ref{sec:pinta_desc}, we have chosen two observations without significant interstellar scintillation in order to show a fair comparison between \gptool{} and \rficlean{}.

Figure \ref{fig:profile_comparison} shows the gain in profile SNR for both datasets while using RFI mitigation.
{Nevertheless, it should be noted that the SNRs for J2124$-$3358 reported by \texttt{pdmp} may be inaccurate due to its large duty cycle.
This does not affect our comparison between the RFI mitigated and non-RFI mitigated datasets as it is clear from the bottom panel of Figure \ref{fig:profile_comparison}(b) that the RFI mitigated profiles agree with each other better than with the non-RFI mitigated profile, indicating a reduction in the noise level. }

\section{Timing of PSR J1909--3744}
\label{sec:J1909_timing}

{ In this section, we demonstrate the capability of \pinta{} to generate profiles from which high-precision TOAs can be derived. 
We use PSR J1909$-$3744 as an example for this purpose.}

\begin{figure}
    \centering
    \includegraphics[scale=0.50]{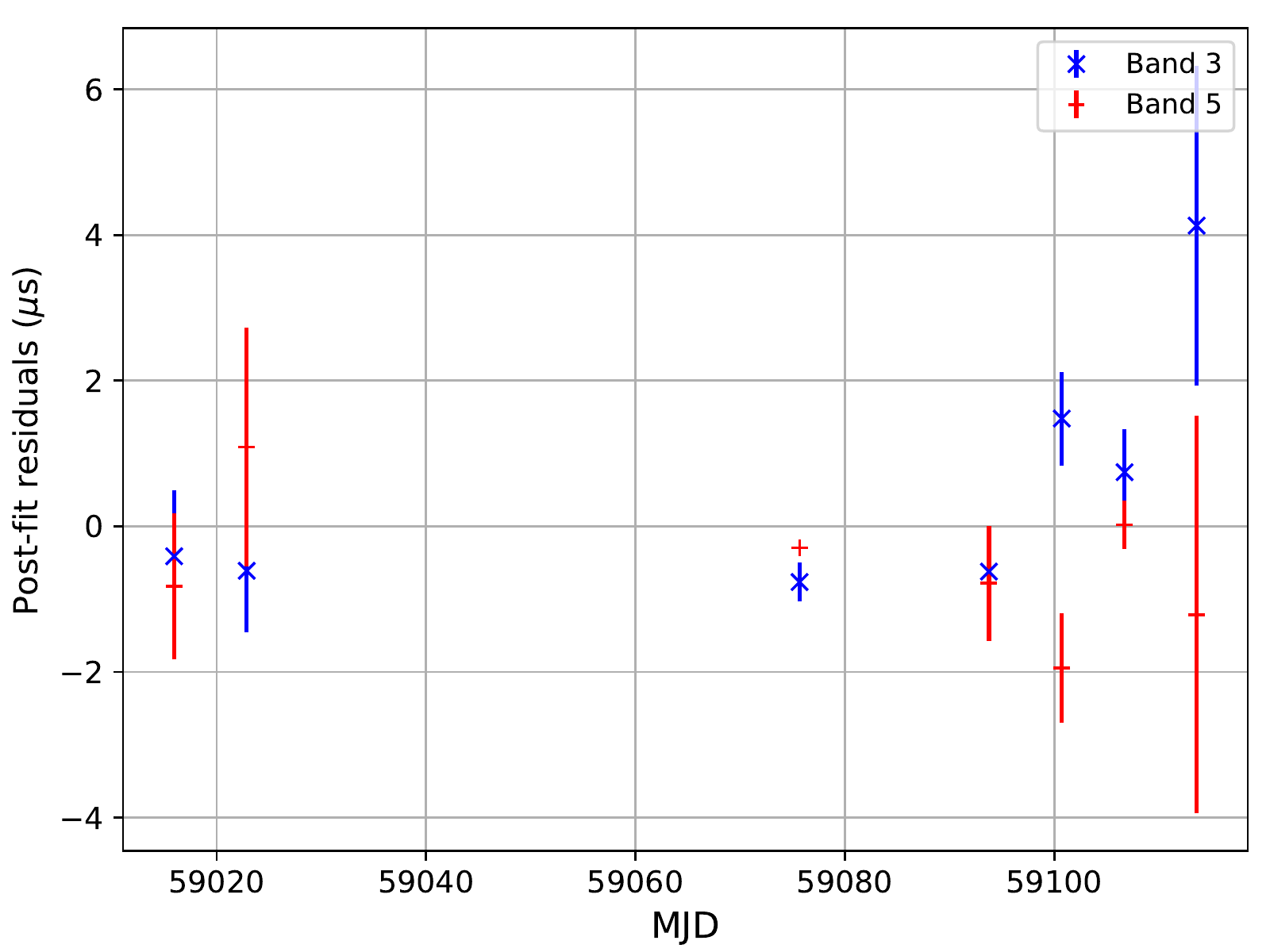}
    \caption{The timing residuals for PSR J1909$-$3744 generated using uGMRT observations processed with \pinta{}.
    Band 3 is 300-500 MHz and Band 5 is 1260-1460 MHz.
    We used the ephemeris available in the NANOGrav 12.5 year dataset, and after changing the PEPOCH and DMEPOCH to MJD 59050, we fitted for F0, F1 and DM.
    The fit parameters are listed in Table \ref{tab:J1909_fit}.
    The timing residuals have an RMS of 1.46 $\mu$s.
    }
    \label{fig:J1909_fit}
\end{figure}

{ The data presented in this section were obtained as part of the InPTA campaign from April 2020 to October 2020 with a cadence of $\sim$15 days.
The observations were carried out by splitting the 30 uGMRT antennas into two phased subarrays, where the innermost 8 antennas were used in Band 3 (300--500 MHz) and 16 of the outer antennas were used in Band 5 (1260--1460MHz).}
{ The pulsar was observed simultaneously in both bands in each epoch, with 200 MHz bandwidth and 1024 frequency channels in each band.}
{ The Band 3  data were coherently dedispersed to the known DM of the pulsar, and were recorded at 20.48 $\mu$s sampling time,  whereas Band 5 data were} obtained using the PA mode with a sampling time of 40.96  $\mu$s.
The data were processed using \pinta{}, and the TOAs were extracted from the resulting \texttt{Timer} archives using \texttt{PSRCHIVE} after time and frequency collapsing the folded profiles.
The resulting TOAs were fit using \texttt{TEMPO2} \citep{Hobbs2006} using the pulsar ephemeris available in the NANOGrav 12.5 year dataset \citep{Alam2020a}, as our data span is too short to provide a reliable timing solution.
{Post-fit residuals after fitting for pulsar rotational parameters (F0, F1), and DM are plotted in Figure \ref{fig:J1909_fit}.
We do not use any time offsets between the two bands as such offsets were corrected in GWB software since April 2020 based on results mentioned in Section \ref{sec:frequency:expt}. 
The corresponding pre-fit and post-fit parameters, along with the  RMS timing residual values  are listed in Table \ref{tab:J1909_fit}. }
A more thorough timing solution of this data using frequency-resolved TOAs, DM corrections and rigorous noise analysis will be published elsewhere.
    
From Table \ref{tab:J1909_fit}, we note that the uGMRT observations processed using \pinta{} are able to produce an RMS post-fit timing residuals of 1.46 $\mu$s. 
This demonstrates that the data products produced using \pinta{} can indeed be used for high-precision timing applications such as PTAs.
We expect to further reduce the RMS timing residuals after applying DM corrections, which are discussed elsewhere {\citep{Krishnakumar2021}}.

\begin{table*}
\centering
\begin{tabular}{lll}
\hline
\textbf{Parameter}                   & \textbf{Post-fit value} & \textbf{Post-fit uncertainty} \\ \hline
F0 (Hz) \dotfill    & 339.315691914442   &       1.2e-12  \\ 
F1 (s$^{-2}$) \dotfill  &  -1.52e-15   &  2.5e-17  \\ 
DM (pc/cm$^3$) \dotfill &  10.39090 & 0.00001  \\ \hline
RMS Residuals ($\mu$s) \dotfill &  1.46    &  \\ \hline
\end{tabular}

\caption{The fit parameters for PSR J1909$-$3744 generated using uGMRT observations processed with \pinta{}.
We used the ephemeris available in NANOGrav 12.5 dataset, and after changing the PEPOCH and DMEPOCH to MJD 59050, and fitted for F0, F1 and DM. 
The timing residuals are plotted in  Figure \ref{fig:J1909_fit}.}
\label{tab:J1909_fit}
\end{table*}


\section{Summary and Discussion}
\label{sec:summary}

We have developed a pipeline to reduce uGMRT pulsar timing raw data for the InPTA experiment, named \pinta{}, which reduces the raw data input to RFI-mitigated folded profile archives.
Since the uGMRT raw data input does not contain any metadata such as the observation settings, they are provided to the pipeline via an ASCII input file named \texttt{pipeline.in}, whose contents are summarized in Table \ref{tab:pipeline.in_cols}.
\pinta{} performs RFI mitigation using two different packages, namely \gptool{} and \rficlean{}, running them in two different branches which produce two different output archives.
\pinta{} provides various command line options to control how these two branches are run, and these are summarized in Table \ref{tab:cmd_options}.

It is crucial to use the correct interpretation of the observatory frequency settings while performing the data reduction.
We performed validation and calibration experiments using GPs from the Crab pulsar and single pulses from the bright pulsar J0332+5434 to ensure that our interpretation of the observation frequency for IA, PA and CDPA pipelines of uGMRT matches what is given in equations (\ref{eq:Fpa}) and (\ref{eq:Fcdp}). 
This experiment also allowed us to measure the instrumental delays between IA, PA and CDPA pipelines of uGMRT, which are consistent with the instrumental delays expected from engineering considerations.

To characterize the computational performance of \pinta{}, we conducted a number of tests using different datasets. 
These tests showed that the net observe-to-reduce time ratio of \pinta{} is approximately 2, while the observe-to-time ratio of individual branches is less than 1.5.
These results lead us to strive to achieve real-time observe-to-time ratio by employing parallelization techniques to the pipeline.
We also conducted tests to investigate the RFI mitigation efficacy of \pinta{} on the same datasets, the results of which are shown in Figure \ref{fig:datagain}. 
We observe that the RFI mitigation gains seen in different datasets, having different RFI characteristics, vary significantly as expected, with some datasets yielding up to $\sim 10\%$ gain after RFI mitigation.
 We also demonstrate improvements in the significance of pulse profiles by using the different RFI mitigation paths in \pinta{}, which further advocates their importance in the pipeline.
These results substantiate the addition of RFI mitigation tools in \pinta{}.
{To demonstrate the ability of \pinta{} to generate data products from which high-precision TOAs can be derived, we showed the timing of uGMRT observations of PSR J1909$-$3744,  
and we are able to produce timing residuals with RMS of the order of 1 $\mu$s.} 


\section{Future scope}
\label{sec:future}
Our plans for the future development of \pinta{} include the improvement of its computational efficiency to achieve better than real-time performance.
This may be achieved by (a) running the two branches of the pipeline parallelly instead of serially, (b) modifying the \filterbank{} program to use {GPUs} and (c) utilizing the GPU processing option in \dspsr{}.

Similar pipelines for reducing the data obtained using the legacy GMRT and the ORT are also under development, ensuring a high level of compatibility with \pinta{}.
In addition, we plan on developing ``InPTA Data Management System'', a database for tracking metadata associated with the observations and data analysis of the InPTA experiment, which will be tightly integrated with \pinta{} as well as the legacy GMRT and ORT pipelines.

\begin{acknowledgements}
We are grateful to the anonymous referee for a detailed  perusal and constructive feedback on  the manuscript.
We thank the staff of the GMRT who made our observations possible. 
GMRT is run by the National Centre for Radio Astrophysics of the Tata Institute of Fundamental Research. BCJ, YG and AB acknowledge the support of the Department of Atomic Energy, Government of India, under project \# 12-R\&D-TFR-5.02-0700. 
AS, AG and LD acknowledge the support of the Department of Atomic Energy, Government of India, under project \# 12-R\&D-TFR-5.02-0200. MPS acknowledges funding from the European Research Council (ERC) under the European Union's Horizon 2020 research and innovation programme (grant agreement No. 694745).
AC acknowledges the funding received from Department of Science and Technology, Government of India, WOS-A scheme, file no. SR/WOS-A/PM-26/2018.
\end{acknowledgements}


\bibliographystyle{pasa-mnras}
\bibliography{pinta}

\end{document}

%% file: affil.tex
\newcommand{\TIFRAstro}{Department of Astronomy and Astrophysics, Tata Institute of Fundamental Research, Dr. Homi Bhabha Road, Mumbai 400005, Maharashtra, India}
\newcommand{\ASTRON}{ASTRON, the Netherlands Institute for Radio Astronomy, Postbus 2, 7990 AA, Dwingeloo, The Netherlands}
\newcommand{\NCRA}{National Centre for Radio Astrophysics, Tata Institute of Fundamental Research, Ganeshkhind, Pune 411007, Maharashtra, India}
\newcommand{\RRI}{Raman Research Institute, Bengaluru 560080, Karnataka, India}
\newcommand{\IITHPhys}{Department of Physics, Indian Institute of Technology Hyderabad, Kandi, Telangana 502285, India}
\newcommand{\BITSHydPhys}{Department of Physics, 
BITS Pilani Hyderabad Campus, 
Hyderabad 500078, Telangana, India}
\newcommand{\IISERTvm}{Indian Institute of Science Education and Research Thiruvananthapuram, Vithura,  Kerala 695551, India}
\newcommand{\IITDPhys}{Department of Physics, Indian Institute of Technology Delhi, New Delhi-110016, India}
\newcommand{\IITRPhys}{Department of Physics, Indian Institute of Technology Roorkee, Roorkee 247667, Uttarakhand, India}
\newcommand{\IMSc}{The Institute of Mathematical Sciences,  C. I. T. Campus, Tharamani, Chennai 600113, Tamil Nadu, India}
\newcommand{\Jodrell}{Jodrell Bank Centre for Astrophysics, University of Manchester, Oxford Road, Manchester, M13 9PL, UK}
\newcommand{\HBNI}{Homi Bhabha National Institute, Training School Complex, Anushakti Nagar, Mumbai 400094, Maharashtra, India}
\newcommand{\UTRGVPhysAstro}{Department of Physics and Astronomy, University of Texas, Rio Grande Valley, Brownsville, TX 78520, USA}
\newcommand{\CaltechCahill}{Cahill Center for Astrophysics, California Institute of Technology, 1200 E. California Blvd. Pasadena, CA 91125, USA}
\newcommand{\UBielefeldPhys}{Fakult{\"a}t f{\"u}r Physik, Universit{\"a}t Bielefeld, Postfach 100131, D-33501 Bielefeld, Germany}
\newcommand{\Arecibo}{Arecibo Observatory, University of Central Florida, Arecibo, FL 32816, USA}
\newcommand{\McGillSpace}{McGill Space Institute, McGill University, 3550 University Street, Montr{\'e}al, QC H3A 2A7, Canada}